\documentclass[prd,superscriptaddress]{revtex4}
\usepackage{hyperref}
\hypersetup{
  colorlinks=true,        
  linkcolor=blue,         
  citecolor=cyan,         
}

\usepackage{graphicx}
\usepackage{dcolumn}
\usepackage{bm}
\usepackage{color}
\usepackage{epsfig}
\usepackage{enumitem}
\usepackage{amsmath}
\usepackage{amssymb}
\usepackage{subcaption}

\begin{document}


\title{Curved spacetime as a dispersive multiferroic medium for an electromagnetic wave: \\ polarization and magnetization vectors in the Schwarzschild spacetime}

\author{Bobur Turimov}
\email{bturimov@astrin.uz}
\affiliation{School of Applied Mathematics, New Uzbekistan University, Mustaqillik Avenue 54, Tashkent 100007, Uzbekistan}
\affiliation{Akfa University, Milliy Bog' Street 264, Tashkent 111221, Uzbekistan}\affiliation{Ulugh Beg Astronomical Institute, Astronomy St 33, Tashkent 100052, Uzbekistan}\affiliation{Department of Civil System Engineering, Ajou University in Tashkent, Asalobod St. 113, Tashkent 100204, Uzbekistan}

\author{Igor Smolyaninov}
\email{igsmoly@gmail.com}
\affiliation{ECE Department, University of Maryland, College Park, Maryland 20742, USA} 

\begin{abstract}
We study one of the interesting properties of the electromagnetic wave propagation in the curved Schwarzschild background spacetime in the framework of general relativity (GR). The electromagnetic wave equation has been derived from vacuum general relativistic Maxwell's equations. It is shown that the solutions for the electromagnetic field can be expanded in the spherical harmonic functions and all components of the electromagnetic fields can be expressed in terms of two radial profile functions. These radial profile functions can be expressed in terms of the confluent Heun function. The calculated behavior of the electric and magnetic susceptibilities near the event horizon appears to be similar to the susceptibilities of multiferroic materials near phase transition. The Curie temperature of this phase transition appears to coincide with the Hawking temperature.
\end{abstract}


\maketitle

\section{Introduction}

All gravitational compact objects in the Universe emit, absorb, reflect and transmit  electromagnetic radiation which allows an observer to get information about the composition, temperature, density, age, motion, distance, and other chemical and physical quantities of such objects. From this point of view, studying the propagation of the electromagnetic wave in the curved space is becoming extremely important and interesting topic of modern astrophysics and cosmology, in particular, after recent observation of the first images of the supermassive black hole (SMBH) candidates located in the centre of galaxy M87 \cite{EHT1,Akiyama19II} and galaxy Sgr A$^*$ \cite{EHT2022ApJ1,EHT2022ApJ2} by the Event Horizon Telescope (EHT) collaboration. There have been also hugely important events of detection of the gravitational waves from the binary systems, in particular, black hole-black hole binary \cite{Abbott2016PRL} and neutron star-neutron star binary \cite{GW170817PhysRevLett} by LIGO/Virgo scientific collaborations. 

It was shown in Ref. \cite{Balazs1958PR} that the gravitational field of a rotating compact object can rotate the direction of the polarization vector of an electromagnetic wave passing in its field. Influence of gravitation on propagation of the electromagnetic wave is studied in \cite{Mashhoon1975PRD}. The diffraction of the electromagnetic wave \cite{Elster1980ASS}, scattering of electromagnetic wave \cite{Mashhoon1973PRD,Crispino2009PRL}, and radiation in the Schwarzschild spacetime \cite{Johnston1973PRL,Tipler1975ApJ,Folacci2018PRD} have been studied. Quasi-normal modes (QNM) of the Schwarzschild black hole were studied in~\cite{Chandrasekhar1975}. 

The vacuum solution for the electromagnetic fields outside a magnetized sphere in the Newtonian framework \cite{Deutsch1955,Ruffini73} and its general-relativistic correction ~\cite{Giznburg1964,Wasserman83,Turimov2021PRD} have been investigated, while the effects of alternative theories of the electromagnetic fields around compact objects have been studied in Refs.~\cite{Ahmedov08,Turimov18}. One of the simplest models of energy loss due to the magneto-dipole radiation from the rotating relativistic star has been studied in \cite{Goldreich69,Thompson04}. A realistic magnetosphere model of the magnetized star due to the light curve of pulsars has been investigated in~\cite{Lockhart19}. The structure of the pulsar magnetosphere, related astrophysical processes and high energy particle acceleration mechanisms have been widely studied, see e.g.~\cite{Mestel71,Arons79,Rayimbaev19a}.  An analytical estimation for the magneto-dipole radiation and oscillations of a highly magnetized relativistic star have also been studied \cite{Rezzolla2016}. The time dependence of dipole magnetic field~\cite{Geppert00,Zanotti2002} and multipole magnetic field~\cite{Mitra1999,Konar1999a} of a magnetized neutron star have been investigated. The magnetic field decay through Hall drift in stellar crusts have been studied in~\cite{Kojima2012MNRAS}. 

Controlling the evolution of light through the space curvature of the medium in the framework of general relativity has been studied in Ref. \cite{Bekenstein2017NaPho} , in particular, they studied trajectory of light in a paraboloid structure inspired by the Schwarzschild metric describing the spacetime around a black hole. It has been shown that this method allows calculations of light-ray trajectories, as well as determination of the diffraction properties and the phase and group velocities of wavepackets propagating within the curved-space structure. The propagation of a light wave in curved thin elastic waveguides, where curvature is shown to be equivalent to a spatially modulated refractive index, has been investigated in \cite{Mazzotti2022}. Introducing topological phases in photonic lattices in curved space \cite{Lustig2017PRA}, it is shown that the curvature of spacetime can induce topological edge states and topological phase transitions in waveguiding layer covering the surface of a three-dimensional body. In Ref. \cite{Pineault77} polarization vector of light in the spacetime of a rotating black hole has been studied, and it is mentioned that polarization occurs due to pure gravitational field and rotation of a spacetime. In particular, it was shown by several authors that the index of refraction of a curved space around a black hole is 
\begin{align}
n=\frac{1}{\sqrt{1-\frac{2M}{r}}}\ .
\end{align}

In this paper, we investigate the polarization  and magnetization vectors in the curved space, in particular in the Schwarzschild spacetime. We demonstrate that the calculated behavior of the electric and magnetic susceptibilities near the event horizon is similar to the susceptibilities of multiferroic materials near phase transition. The reinterpretation of this behavior as a critical phenomenon is novel and very interesting, especially in view of the fact that the Curie and the Hawking temperatures coincide in this novel physical picture.  

The paper is organized as follows. In Sect.~\ref{sec1}, we discuss the derivation of the expressions for the polarization and magnetization vector for the propagating light-ray in the Schwarzschild spacetime. The analogy between the calculated behavior of the electric and magnetic susceptibilities near the event horizon and the susceptibilities of multiferroic materials near phase transition is described. In Sect.~\ref{sec2}, we present our results on light propagation in the Schwarzschild spacetime. Finally, we summarize the obtained results in Section~\ref{sec3}. Throughout the paper, we use a space-like signature $(-,+,+,+)$, a~system of units in which $G=c=1$. Greek indices are taken to run from $0$ to $3$, while Latin indices from $1$ to $3$.

\section{Polarization and magnetization vectors and electric and magnetic susceptibilities:  Schwarzschild spacetime as a dispersive multiferroic medium}\label{sec1}

It is well known that the polarization and the magnetization vectors of a medium are determined as
\begin{align}
{\mathbf P}={\mathbf D}-{\mathbf E}\ ,\qquad {\mathbf M}={\mathbf B}-{\mathbf H}\ , \end{align}
and relations between the vectors $\mathbf{D}$ and $\mathbf{E}$, as well as vectors $\mathbf{B}$ and $\mathbf{H}$ in the macroscopic electrodynamics are defined as
\begin{align}
{\mathbf D}=\epsilon{\mathbf E}\ ,\qquad {\mathbf B}=\mu{\mathbf H}\ .     
\end{align}
where $\epsilon$ and $\mu$ are the effective electric permittivity and magnetic permeability of the spacetime defined as $\epsilon=\mu=1/\sqrt{-g_{tt}}$. On the other hand, the effective polarization vector is proportional to the electric field, while the effective magnetization vector is proportional to the magnetic field, or
\begin{align}
{\mathbf P}=\chi{\mathbf E} \ , \qquad {\mathbf M}=\chi'{\mathbf H}\ ,   
\end{align}
where $\chi$ is the electric susceptibility  and $\chi'$ is the magnetic susceptibility. Taking into account all these facts, one can get 
\begin{align}
\epsilon=\mu=\frac{1}{\sqrt{1-\frac{2M}{r}}}\ ,\quad\to \quad \chi=\chi'=\frac{1}{\sqrt{1-\frac{2M}{r}}}-1\ .
\end{align}

Note that the effective permeability and permittivity in the Schwarzschild metric are isotropic. The respective tensors are diagonal, and all the diagonal elements are the same.

Using the explicit solution for the electromagnetic field (see, e.g. App. \ref{app}), the components of the polarization vector in the Schwarzschild spacetime are
\begin{align}
&P_{\hat r}=\frac{e^{-i\omega t}}{r}\left(\frac{1}{\sqrt{f}}-1\right)V_{\ell}Y_{\ell m}\ ,
\\
&P_{\hat\theta}=\frac{e^{-i\omega t}(1-\sqrt{f})}{\ell(\ell+1)}\left[D_rV_{\ell}\partial_\theta Y_{\ell m}+\frac{i\omega}{f\sin\theta}U_{\ell}\partial_\phi Y_{\ell m}\right]\ ,
\\
&P_{\hat\phi}=\frac{e^{-i\omega t}(1-\sqrt{f})}{\ell(\ell+1)}\left[D_rV_{\ell}\frac{1}{\sin\theta}\partial_\phi Y_{\ell m}-\frac{i\omega}{f}U_{\ell}\partial_\theta Y_{\ell m}\right]\ ,
\end{align}
and the components of magnetization vector are
\begin{align}
&M_{\hat r}=\frac{e^{-i\omega t}}{r}(1-\sqrt{f})U_{\ell}Y_{\ell m}\ ,
\\
&M_{\hat\theta}=\frac{e^{-i\omega t}\sqrt{f}(1-\sqrt{f})}{\ell(\ell+1)}\left[D_rU_{\ell}\partial_\theta Y_{\ell m}-\frac{i\omega}{f\sin\theta}V_{\ell}\partial_\phi Y_{\ell m}\right]\ ,
\\
&M_{\hat\phi}=\frac{e^{-i\omega t}\sqrt{f}(1-\sqrt{f})}{\ell(\ell+1)}\left[D_rU_{\ell}\frac{1}{\sin\theta}\partial_\phi Y_{\ell m}+\frac{i\omega}{f}V_{\ell}\partial_\theta Y_{\ell m}\right]\ .
\end{align}

This representation is important as a particular example of a situation in which all components of the electromagnetic field and its vector potential may be found in analytical form in the Schwarzschild metric.

Here we will consider transverse magnetic waves (TM mode) characterized by the fact that the magnetic vector (${\bold B}$) is always perpendicular to the direction of propagation. In this case the components of polarization and magnetization vectors are
\begin{align}
&P_{\hat r}=\frac{e^{-i\omega t}}{r}\left(\frac{1}{\sqrt{f}}-1\right)V_{\ell}Y_{\ell m}\ ,
\\
&P_{\hat\theta}=\frac{e^{-i\omega t}(1-\sqrt{f})}{\ell(\ell+1)}D_rV_{\ell}\partial_\theta Y_{\ell m}\ ,
\\
&P_{\hat\phi}=\frac{e^{-i\omega t}(1-\sqrt{f})}{\ell(\ell+1)}D_rV_{\ell}\frac{1}{\sin\theta}\partial_\phi Y_{\ell m}\ ,\\
&M_{\hat r}=0\ ,
\\
&M_{\hat\theta}=-\frac{e^{-i\omega t}}{\ell(\ell+1)}\left(\frac{1}{\sqrt{f}}-1\right)\frac{i\omega}{\sin\theta}V_{\ell}\partial_\phi Y_{\ell m}\ ,
\\
&M_{\hat\phi}=\frac{e^{-i\omega t}}{\ell(\ell+1)}\left(\frac{1}{\sqrt{f}}-1\right)i\omega V_{\ell}\partial_\theta Y_{\ell m}\ ,
\end{align}
while in the case of transverse electric waves (TE mode) which is characterized by the fact that the electric vector (${\bold E}$) is always perpendicular to the direction of propagation, one can get
\begin{align}
&P_{\hat r}=0\ ,
\\
&P_{\hat\theta}=\frac{e^{-i\omega t}(1-\sqrt{f})}{\ell(\ell+1)}\frac{i\omega}{f\sin\theta}U_{\ell}\partial_\phi Y_{\ell m}\ ,
\\
&P_{\hat\phi}=-\frac{e^{-i\omega t}(1-\sqrt{f})}{\ell(\ell+1)}\frac{i\omega}{f}U_{\ell}\partial_\theta Y_{\ell m}\ ,
\\
&M_{\hat r}=\frac{e^{-i\omega t}}{r}(1-\sqrt{f})U_{\ell}Y_{\ell m}\ ,
\\
&M_{\hat\theta}=\frac{e^{-i\omega t}(\sqrt{f}-f)}{\ell(\ell+1)}D_rU_{\ell}\partial_\theta Y_{\ell m}\ ,
\\
&M_{\hat\phi}=\frac{e^{-i\omega t}(\sqrt{f}-f)}{\ell(\ell+1)}D_rU_{\ell}\frac{1}{\sin\theta}\partial_\phi Y_{\ell m}\ .
\end{align}

\begin{figure}[t]
\includegraphics[width=\hsize]{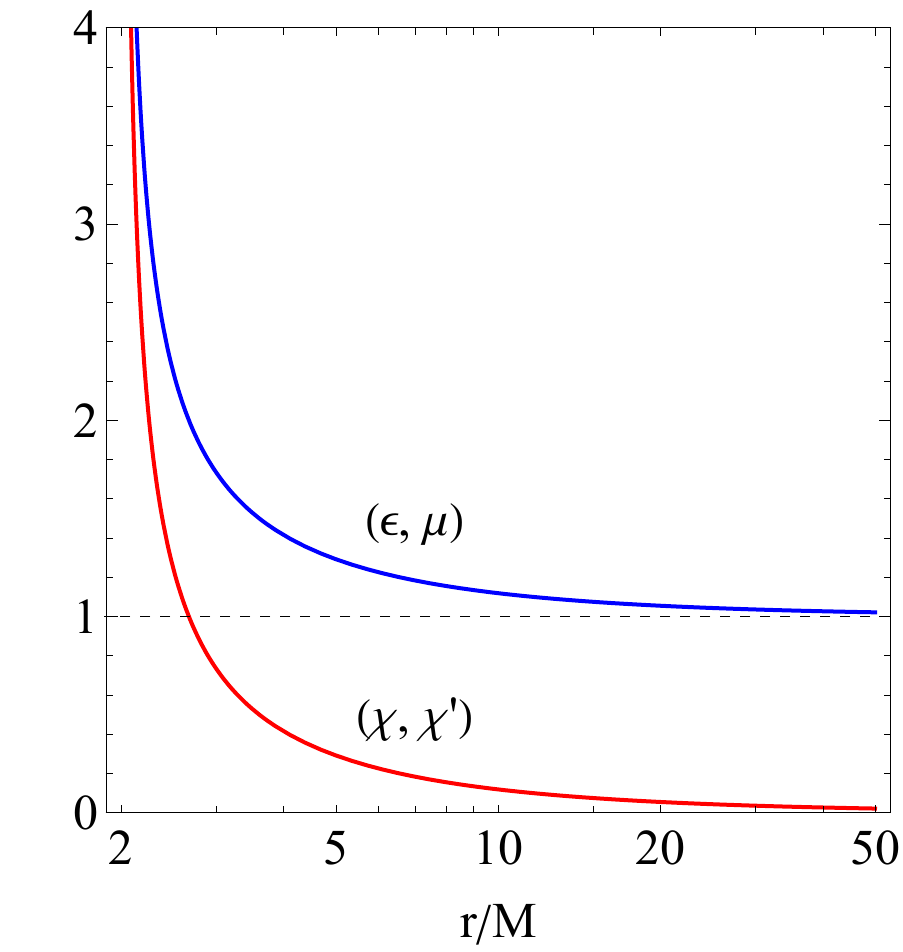}\caption{The radial dependence of the electric susceptibility and magnetic susceptibility.\label{fig}}
\end{figure}

We should note that the divergence of the electric and magnetic susceptibilities of vacuum near the event horizon looks very similar to the properties of a multiferroic material \cite{Spaldin20,Fiebig16,Dong15} near its phase transition. Indeed, due to the Unruh effect, a near-horizon observer must see the electromagnetic field excited at a local temperature 
\begin{align}
T_{\rm loc}=\frac{1}{4\pi \sqrt{2Mr(1-\frac{2M}{r})}}\ 
\end{align}
Therefore, after simple algebraic transformations, the derived expressions for the electric and magnetic susceptibilities may be recast as
\begin{align}
\chi\sim\frac{1}{(T-T_H)^\gamma}\ ,
\end{align}
where $T_H=\frac{1}{8\pi M}$ is the Hawking temperature,
\begin{align}
T=\frac{1}{4\pi \sqrt{2Mr}}\ 
\end{align}
is the local temperature experienced by an observer near horizon at radius $r$ redshifted to infinity, and $\gamma=1/2$  is the critical exponent. It is evident that this expression looks similar to the Curie-Weiss law in multiferroics. 

Below some critical temperature $T_C$, multiferroic materials simultaneously exhibit ferromagnetic and ferroelectric properties. In some cases Tc may be equal to zero, so that a quantum critical point is observed. Above $T_C$, very near the critical temperature these materials exhibit simultaneously divergent paramagnetism and paraelectricity. While they have zero net dipole moments, their electric and magnetic susceptibilities diverge:

\begin{align}
\chi=\frac{C}{T-T_C}\ ,
\end{align}
where $C$ is the material-specific Curie-Weiss constant (for typical multiferroic materials in three spatial dimensions the critical exponent equals $\gamma=1$). Below $T_C$ these materials exhibit simultaneous ferromagnetism and ferroelectricity, and their susceptibilities also diverge near the critical temperature. We should also mention the possibility of reverse "re-entrant" behavior in semiconductors doped with magnetic impurities, in which the paramagnetic state exists at low temperatures, while heating the material above $T_C$ leads to establishment of the ferromagnetic order \cite{Calderon07}. Re-entrant ferroelectricity was also observed in multiferroic Fe-substituted MnWO4 \cite{Chaudhury09}.  

It is important to note that the expected critical exponent in a 4D Schwarzschild spacetime does not need to be 1. In general, different theories, such as mean field theory in various dimensions, the two-dimensional Ising model, etc. produce different values of the critical exponents, and these exponents do not necessarily coincide with the experimentally measured values. Therefore, the obtained 1/2 value of the critical exponent should not be considered as not perfect.  

Typically, the multiferroic materials also exhibit ferroelasticity \cite{Salje12}, which implies that elastic deformations of these materials strongly affect their magnetic and electric properties \cite{Zhang21}. For example, such materials as FeZrB(Cu) spin glasses exhibit very pronounced dependencies of their Curie temperature $T_C$ on tensile stress \cite{Barandiaran96}. In a similar fashion, we may interpret Fig.\ref{fig} as an evidence of a phase transition phenomenon in which a deformation of Minkowski spacetime due to gravity leads to the onset of its multiferroic properties.

Since the divergent magnetic and electric susceptibilities of the Minkowski vacuum are identical, we should classify the physical vacuum as type-II multiferroic material (based on the clasification developed by Khomskii \cite{Khomskii09}). In some type II multiferroics magnetic ordering breaks the inversion symmetry and directly causes the ferroelectricity, so that the ordering temperatures for the two phenomena are identical. The typical example of such materials is TbMnO3 \cite{Kimura03}, in which a non-centrosymmetric magnetic spiral structure accompanied by a ferroelectric polarization sets in at $T_C=28$ K. An opposite situation was observed in some Mott insulating charge-transfer salts \cite{Lang14}, where a charge-ordering transition to a polar ferroelectric state may drive magnetic ordering.

We should also note that reinterpretation of a quantum black hole using the language of modern condensed matter physics became a very active research field recently \cite{Dvali14,Stephens01}. For example, Dvali $et$ $al.$ \cite{Dvali14} proposed that a black hole may be understood as a graviton Bose-Einstein condensate at the critical point of a quantum phase transition, which is somewhat similar to phase transitions observed in cold atoms. In agreement with our multiferroic analogy, it may be argued that graviton condensation is also somewhat similar to the ferroelastic phase transition, in which spontaneous strain arises in a material in response to an applied stress. In another example, Stephens $et$ $al.$ suggested that the black hole system shares similarities with the defect-mediated Kosterlitz-Thouless transition \cite{Kosterlitz72} in condensed matter. Thus, our interpretation of Fig.\ref{fig} as a multiferroic phase transition makes perfect sense within the context of this contemporary line of thought.  

We must also point out that regardless of the coordinate choice, the effective permittivity and permeability of the spacetime metric describing a spherically symmetric distribution of mass cannot be eliminated everywhere. In this sense, the polarization and the magnetization of spacetime should be treated as essential physical objects. In some way, these properties of the physical spacetime are very similar to the properties of any real electromagnetic medium. For example, it is well established in Transformation Optics that by transforming the macroscopic Maxwell equation into some curvilinear coordinate system, a local value of the effective permittivity and permeability may be always made equal to 1 in any given location inside the medium. However, this fact does not make such a medium any less real. 

Based on the above discussion, and by combining Eqs.(5) and (27), we may conclude that light propagation in the Schwarzschild spacetime may be emulated by engineering a hot spot inside a multiferroic medium in which the temperature distribution near horizon behaves as

\begin{align}
T=T_C+C\sqrt{1-\frac{2M}{r}}\ ,
\end{align}
where $2M$ now defines the effective Schwarzschild radius of an effective electromagnetic black hole inside a multiferroic medium. The scale of $2M$ should be large enough to ensure the validity of macroscopic electrodynamic description of the medium, which means that it can span a range from a few nanometers to macroscopic dimensions. Engineering such a temperature distribution using localized heat sources (or localized absorbers of external radiation) is a straightforward task for modern metamaterial engineering. Moreover, local lattice distortions under the influence of external radiation and heat (which affect the local magnitude of the Curie constant $C$) may also be incorporated into such model.  

In the next Section we will discuss some interesting aspects of light propagation around both astrophysical and emulated Schwarzschild spacetime. However, we should emphasize the difference in physical mechanisms involved in these two different situations. In the astrophysical case, relativistic solutions for the $E$- and $B$-field are the reason for curvilinear light ray trajectories. In the second one, the magnetization and polarization of matter lead to the observable optical effects. 

\section{Light propagation in the Schwarzschild spacetime}\label{sec2}

As one can see from the equations derived in Section 2, it is sufficient to find the profile functions $U_\ell$ and $V_\ell$ in order to produce all components of the electromagnetic field and its vector potential. Now inserting the expressions (\ref{sol1}) and \eqref{sol2} into (\ref{E2}) and \eqref{B2}, the radial equation is expressed as
\begin{align}\label{rad}
\left[f\partial_r \left(f\partial_r\right)+\omega^2-f\frac{\ell(\ell+1)}{r^2}\right]\Big[rU_{\ell}(r), rV_{\ell}(r)\Big]=0\ ,
\end{align}
which is also known as the Regge-Wheeler equation for the function $rV_\ell(r)$. Notice that the same equation can be obtained for the function $U_\ell(r)$. It is well known that in a flat space, (i.e. $f=1$), the solution $V_\ell(r)$ for the electromagnetic wave equation is expressed in terms of the spherical Bessel functions of the first and second kind, $j_\ell(\omega r)$ and $y_\ell(\omega r)$ which describe standing waves with regularity and singularity at the origin; or the spherical Bessel function of third kind, also known as Hankel functions, $h_\ell^\pm(\omega r)$ which describe travelling waves with a source at the origin where they have a singularity, see e.g. \cite{Rezzolla04}. Here the upper indices ``$\pm$'' denote outgoing and ingoing waves.

\begin{figure}
\includegraphics[width=\hsize]{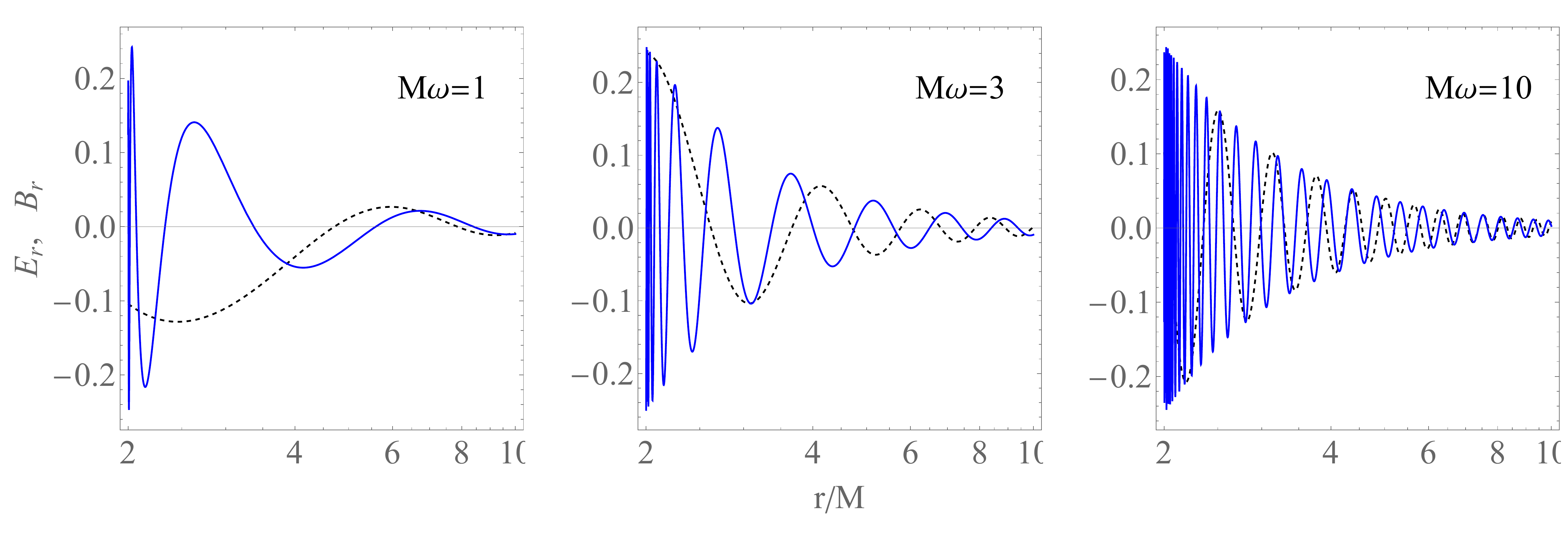}\caption{The radial dependence of the monopole electromagnetic wave for different angular frequency in the Schwarzschild spacetime (solid blue line) and in a flat space (dashed black line).\label{fig1}}
\end{figure}

In general, finding the solution of equation (\ref{rad}) is not a simple task. However, for a particular case, when $\ell=0$, an analytical solution of the wave equation may be found as $\sim e^{\pm i\Omega r}$ with $\Omega=\omega[1+(1-f)\ln f]$, which is responsible for a monopole solution describing a time dependent monopole point-like charge. However, from the physical point of view the magnetic monopole solution is meaningless. One has to emphasise that in this case the components of the vector potential are divergent. Therefore, monopole solutions can be directly found by solving equations (\ref{E2}) and (\ref{B2}), so that $E_{\hat r}, B_{\hat r}\sim r^{-2}e^{\pm\Omega r}$. These solutions are not regular at the origin however, at the horizon they have certain value. Figure \ref{fig1} draws radial dependence of the monopole electromagnetic wave in the Schwarzschild spacetime (solid blue line) and in a flat space (dashed black line) for a different value of the angular frequency. As one can see, there is no difference of the amplitude of the electromagnetic wave. On the other hand, there is a time delay between the propagation of waves in the flat and curved space. In the near zone the field lines in curved space are quite dense in comparison with that in flat space. 

It is worth noticing that for the very small angular frequency i.e. $M\omega\ll 1$, the second term of Eq. (\ref{rad}) may be negligible. In this case, the equation will be quite simple and the stationary solution can be expressed in terms of the special functions, for instance, the reduced Legendre functions of the first and second kind $\sim P_\ell^1(1-r/M)$ and $\sim Q_\ell^1(1-r/M)$ in Refs. \cite{Rezzolla01c,Rezzolla04,Turimov2021PRD}, the hypergeometric function $\sim\,_2F_1(\ell,\ell+2,2(\ell+1),2M/r)$ in \cite{Muslimov1986,Muslimov1986a,Petri2017} and the  Jacobi polynomial $\sim J_{\ell-1}^{(2,0)}(1-r/M)$ (details are to be found in Ref.\cite{Petterson1974PRD}). Notice that those special functions are not fully solutions of the wave equation. However, these solutions are valid when the $\omega^2$ term is very small in comparison with the other terms of Eq. (\ref{rad}), or in the stationary case.  

The numerical solution of the wave equation in the background of Schwarzschild spacetime is presented by number of authors, for instance in the WKB approximation \cite{Iyer87a,Iyer87b,Konoplya03,Turimov2019PRD} and semi-analytical approach has been suggested in \cite{Leaver1985,Schutz1985}. Introducing the tortoise coordinate $r_*=r+2M\ln(r/2M-1)$, the equation (\ref{rad}) reduces to the standard one dimensional Schrodinger-like equation for a particle with energy $\omega^2$ in the potential $\ell(\ell+1)f/r^2$, (See, e.g. \cite{Fabbri1975PRD}). Very similar equations can be easily derived for the spin zero and spin two fields in the scalar and gravitational perturbations. Unlike other perturbations, the stationary point of the given potential in the electromagnetic perturbation is independent of azimuthal number $\ell$, or $r_c=3M$, which is responsible for the peak of the potential (a maximum of potential). The critical value of the angular frequency at the turning is $\omega_c=\sqrt{\ell(\ell+1)}/(3\sqrt{3}M)$. If the angular frequency $\omega$ is a greater than the critical one (i.e. $\omega>\omega_c$) the wave vector will be real, otherwise ($\omega<\omega_c$) it should be imaginary \cite{Fabbri1975PRD}. At the large distance, $r\to\infty$, the solution of Eq. (\ref{rad}) is expressed as $rV_\ell\sim e^{\pm i\omega r}$, which is independent of the orbital number $\ell$. 

\begin{figure}
\includegraphics[width=\hsize]{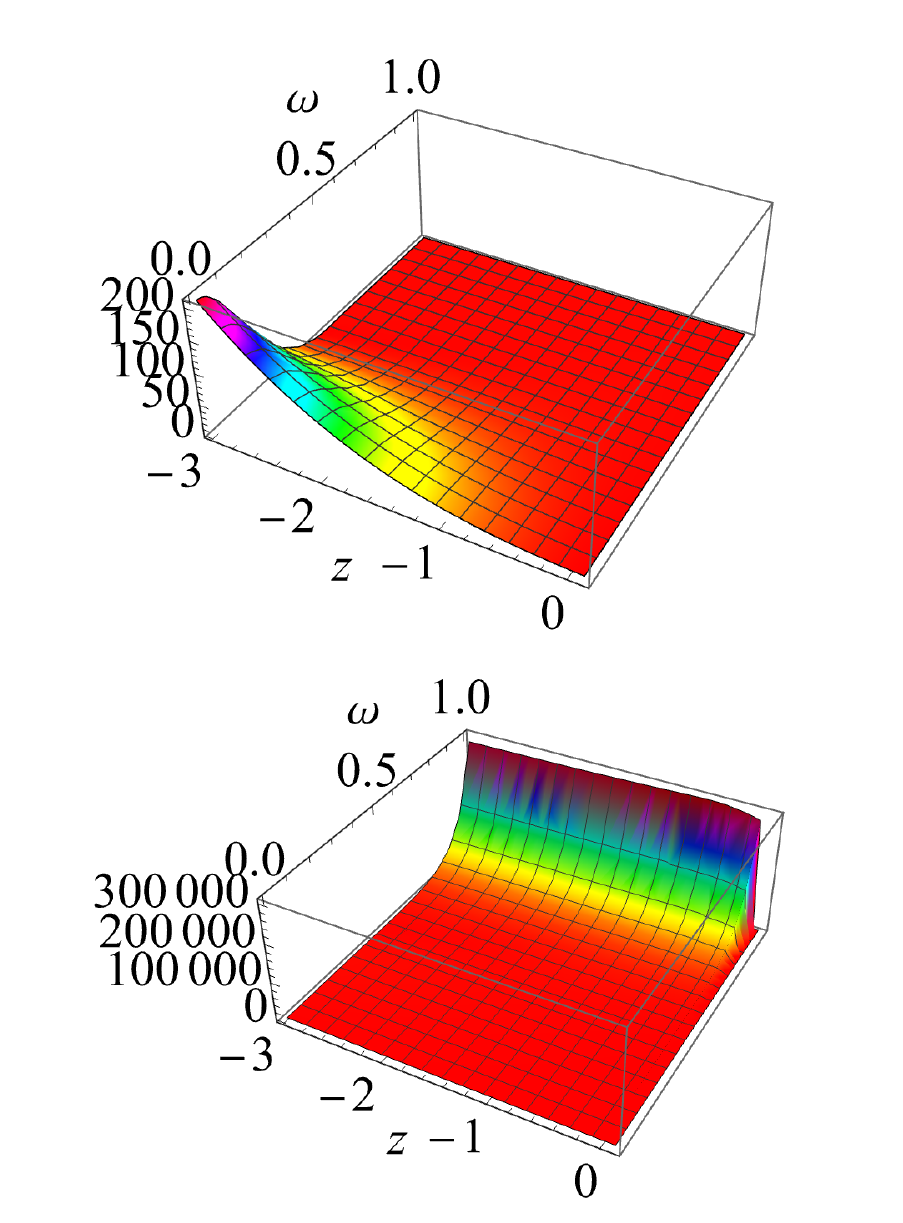}\caption{The confluent Heun function as a function of $z$ and $\omega$.\label{fig2}}
\end{figure}

The main purpose of the paper is finding the exact analytical solution for the radial waves Eq. (\ref{rad}). Hereafter introducing new radial coordinate, $z=1-r/2M$, and redefining the profile function $V_{\ell}(z)=F(z)e^{\epsilon z/2}z^{(\gamma-1)/2}(z-1)^{(\delta-1)/2}$, the radial equation reduces to the well-known confluent Heun equation \cite{Heun1888AM,Fiziev2011PRD,Boyack2011JMP}: 
\begin{align}
F''+\left(\frac{\gamma}{z}+\frac{\delta}{z-1}+\epsilon\right)F'+\frac{\alpha z-q}{z(z-1)}F=0\ ,    
\end{align}
which has three singular points at $z=0$, $z=1$ and $z=\infty$. The solution to above equation is represented by the confluent Heun function: 
\begin{align}\nonumber
F(z)&=C_{1\ell}{\rm HeunC}(q, \alpha, \gamma, \delta, \epsilon, z)\\&+C_{2\ell}z^{1-\gamma}{\rm HeunC}[q+(\gamma-1)(\delta-\epsilon),\alpha-\gamma\epsilon+\epsilon, 2-\gamma, \delta, \epsilon, z]\ ,
\end{align}
where $C_{1\ell}$ and $C_{2\ell}$ are integration constants, and the explicit form of the parameters $\epsilon$, $\gamma$, $\delta$, $\alpha$ and $q$ are listed in Tab \ref{tab}. In general, there are eight different cases for the coefficients. The careful analyses showed that solutions $F(z)$ at some of parameter values are identical and there are only two different cases for all combinations of parameters. These two solutions are responsible for the incoming and outgoing waves, which is illustrated in Fig. \ref{fig2}. Therefore, it is sufficient to use one of solutions of $F(z)$ for the specific form of the parameters. The Wronskian of the solutions is given by $W=e^{-\epsilon z}z^{-\gamma}(z-1)^{-\delta}$. At the horizon of the black hole (i.e. $r=2M$ or $z=0$) the function takes a form $F(0)=C_{1\ell}$. At very small values of the argument, the confluent Heun function reduces to ${\rm HeunC}(q, \alpha, \gamma, \delta, \epsilon, z)\simeq 1-(q/\gamma) z+{\cal O}\left(z^2\right)$. 

\begin{table}
\caption{\label{tab}The explicit form of the parameters of the confluent Heun function are listed.}
\begin{tabular}{ccccc}
 $\delta$ & $\epsilon$ & $\gamma$ & $\alpha$ & $q$\\ \hline
$-1$ & $\pm 4iM\omega$ & $1\mp 4iM\omega$ & $0$ &  $\ell(\ell+1)$ \\
$-1$ & $\pm 4iM\omega$ & $1\pm 4iM\omega$ & $-16M^2\omega^2$ & $\ell(\ell+1)-4M\omega(4M\omega\mp i)$ 
\\
$3$ & $\pm 4iM\omega$ & $1\mp 4iM\omega$ & $\pm 8iM\omega$ & $(\ell-1)(\ell+2)\pm 8iM\omega$ 
\\
$3$ & $\pm 4iM\omega$ & $1\pm 4iM\omega$ & $-8iM\omega(2M\omega\mp i)$ & $(\ell-1)(\ell+2)-4M\omega(4M\omega\mp i)$ \\
\end{tabular}
\end{table}
\unskip

\section{Conclusions}\label{sec3}

In the present research, we have studied the propagation of the electromagnetic field in the Schwarzschild spacetime given by metric function $f=1-2M/r$. One must mention that this method is also a valid form of the metric function. We first explicitly derived the electromagnetic wave equation for the two independent radial profile functions, which represent all the components of the electromagnetic field. We have also presented an analytical expression for the components of the vector potential in terms of the radial profiles, which allows to produce electromagnetic field lines for a proper choice of the integration constants. The exact analytical solution of the wave equation for the monopole electromagnetic field has been obtained. We have discussed solutions for the electromagnetic field in several cases, in particular, for a small frequency range. Finally, we have explicitly showed that the confluent Heun function satisfies the radial wave equations for both profile functions $U_\ell(r)$ and $V_\ell(r)$ for a combination of the specific parameters. It is shown that there are only two independent solutions for each profile functions for the specific values of the parameters, which represent incoming and outgoing waves. In the end we have discussed two different modes of wave propagation known as transverse electric wave (TE) and transverse magnetic wave (TM), which are important for the astrophysical consequences.
We also concluded that the calculated behavior of the electric and magnetic susceptibilities near the event horizon appears to be similar to the susceptibilities of multiferroic materials near phase transition. The reinterpretation of this behavior as a critical phenomenon is novel and very interesting, especially in view of the fact that the Curie and the Hawking temperatures coincide in this novel physical picture.

\acknowledgments{This research is supported by Grants F-FA-2021-510 and MRB-2021-527 of the Uzbekistan Ministry for Innovative Development.}

\appendix

\section{Electromagnetic wave equations \label{app}}

The explicit form of the general relativistic Maxwell equations in the Schwarzschild spacetime (in a geometrized units $G=c=1$) for the component of the electromagnetic field measured by a local observer are~\cite{Turimov2021PRD}:
\begin{align}
\label{max1a}
&\frac{\sqrt{f}}{r}\partial_r\left(r^2E_{\hat r}\right)+\frac{1}{\sin\theta}\left[\partial_\theta\left(\sin\theta E_{\hat \theta}\right)+\partial_\phi E_{\hat \phi}\right] = 0\ , 
\\
\label{max1b}
&\partial_tE_{\hat r}
=\frac{\sqrt{f}}{r\sin\theta}\left[\partial_\theta\left(\sin\theta B_{\hat \phi}\right)-\partial_\phi B_{\hat\theta}\right] \ ,
\\
\label{max1c}
&\partial_tE_{\hat\theta} =\frac{\sqrt{f}}{r\sin\theta}\left[\partial_\theta B_{\hat r}-\sin\theta\partial_r \left(r\sqrt{f}B_{\hat \phi}\right)\right]\ ,
\\
\label{max1d}
&\partial_tE_{\hat\phi} = \frac{\sqrt{f}}{r}\left[\partial_r\left(r\sqrt{f}B_{\hat \theta}\right)-\partial_\theta B_{\hat r}\right]\ ,
\end{align}
and 
\begin{align}
\label{max2a}
&\frac{\sqrt{f}}{r}\partial_r\left(r^2B_{\hat r}\right)+\frac{1}{\sin\theta}\left[\partial_\theta\left(\sin\theta B_{\hat\theta}\right)+\partial_\phi B_{\hat\phi}\right] = 0 \ ,
\\
\label{max2b}
&\partial_tB_{\hat r} = \frac{\sqrt{f}}{r\sin\theta}\left[\partial_\phi E_{\hat\theta} - \partial_\theta\left(\sin\theta E_{\hat \phi}\right)\right] \ ,
\\
\label{max2c}
&\partial_tB_{\hat\theta}
=\frac{\sqrt{f}}{r\sin\theta}\left[\sin\theta\partial_r \left(r\sqrt{f}E_{\hat\phi}\right) - \partial_\phi E_{\hat r}\right]\ ,
\\
\label{max2d}
&\partial_tB_{\hat\phi} =  \frac{\sqrt{f}}{r}\left[\partial_\theta E_{\hat r}-\partial_r\left(r\sqrt{f}E_{\hat\theta}\right)\right]\ , 
\end{align}
where $f=1-2M/r$ is the metric function in the Schwarzschild spacetime parameterized by the black hole mass $M$. The components of the electromagnetic fields $E_{\hat i}$ and $B_{\hat i}$ are measurable quantities by a proper observer ($i=r,\theta,\phi$). Acting by the time derivative operator $\partial_t$ on both sides of the radial equations (\ref{max1b}), (\ref{max2b}) and taking into account other Maxwell equations, the second order radial equations can be obtained as 
\begin{align}\label{E2}
\partial_t^2E_{\hat r} = \frac{f}{r^2}\partial_r \left[f\partial_r\left(r^2E_{\hat r}\right)\right]+\frac{f}{r^2}\Delta_\Omega E_{\hat r}\ , \\\label{B2} \partial_t^2B_{\hat r}=\frac{f}{r^2}\partial_r\left[f\partial_r\left(r^2B_{\hat r}\right)\right]+\frac{f}{r^2}\Delta_\Omega B_{\hat r}\ ,
\end{align}
where $\Delta_\Omega$ is the angular part of the Laplace operator which satisfies the following relation, $\Delta_\Omega Y_{\ell m}=-\ell(\ell+1)Y_{\ell m}$, where $Y_{\ell m}(\theta,\phi)$ are the spherical harmonics with the orbital number $\ell=0, 1, 2,...$ and azimuthal number $|m|\leq\ell$, (see e.g. \cite{Arfken2005}). 

The general solutions of Maxwell's vacuum equations (\ref{max1a})-(\ref{max2d}) for the electromagnetic wave can be expressed as~\cite{Turimov2021PRD}
\begin{align}\label{sol1}
&E_{\hat r}=\frac{e^{-i\omega t}}{r}V_{\ell}Y_{\ell m}\ ,
\\
&E_{\hat\theta}=\frac{e^{-i\omega t}\sqrt{f}}{\ell(\ell+1)}\left[D_rV_{\ell}\partial_\theta Y_{\ell m}+\frac{i\omega}{f\sin\theta}U_{\ell}\partial_\phi Y_{\ell m}\right]\ ,
\\
&E_{\hat\phi}=\frac{e^{-i\omega t}\sqrt{f}}{\ell(\ell+1)}\left[D_rV_{\ell}\frac{1}{\sin\theta}\partial_\phi Y_{\ell m}-\frac{i\omega}{f}U_{\ell}\partial_\theta Y_{\ell m}\right]\ ,
\end{align}
and
\begin{align}\label{sol2}
&B_{\hat r}=\frac{e^{-i\omega t}}{r}U_{\ell}Y_{\ell m}\ ,
\\
&B_{\hat\theta}=\frac{e^{-i\omega t}\sqrt{f}}{\ell(\ell+1)}\left[D_rU_{\ell}\partial_\theta Y_{\ell m}-\frac{i\omega}{f\sin\theta}V_{\ell}\partial_\phi Y_{\ell m}\right]\ ,
\\
&B_{\hat\phi}=\frac{e^{-i\omega t}\sqrt{f}}{\ell(\ell+1)}\left[D_rU_{\ell}\frac{1}{\sin\theta}\partial_\phi Y_{\ell m}+\frac{i\omega}{f}V_{\ell}\partial_\theta Y_{\ell m}\right]\ ,
\end{align}
where $\omega$ is the angular frequency of the electromagnetic wave, $U_{\ell}(r)$, $V_{\ell}(r)$ are, respectively, the profile functions of the electromagnetic wave, also known as the Debye potentials and the differential operator $D_r$ is defined as $D_rU_\ell=r^{-1}\partial_r(rU_\ell)$. Thus, the explicit components of the vector potential of the electromagnetic wave are \cite{Turimov2021PRD}
\begin{align}
&A_t=\frac{e^{-i\omega t}}{\ell(\ell+1)}f\partial_r\left(rV_{\ell}\right)Y_{\ell m}\ ,
\\ 
&A_r=-\frac{e^{-i\omega t}}{\ell(\ell+1)}\frac{i\omega r}{f}V_{\ell}Y_{\ell m}\ , 
\\
&A_\theta=\frac{e^{-i\omega t}}{\ell(\ell+1)}rU_{\ell}\frac{1}{\sin\theta}\partial_\phi Y_{\ell m}\ ,
\\ 
&A_\phi=-\frac{e^{-i\omega t}}{\ell(\ell+1)}rU_{\ell}\sin\theta\partial_\theta Y_{\ell m}\ .
\end{align}   
Notice that the above expressions are valid for the spherical wave, while for a plane wave one gets $A_\theta=0$ and the spherical harmonics $Y_{\ell m}(\theta,\phi)$ should be replaced by the associated Legendre polynomial $P_\ell^m(\cos\theta)$, see e.g. \cite{Arfken2005}.

\bibliography{Ref,gravreferences}
\end{document}